\documentstyle[]{mn}
 
\title[HI Streaming Shocks in NGC4151]{Gas Dynamics in the Barred
Seyfert Galaxy NGC4151 - I. HI Streaming Shocks and Inflow Along the Bar}

\author[C.G. Mundell \& D.L. Shone]
        {C.G. Mundell\footnote{Now at Department of Astronomy,
        University of Maryland, College Park, MD~20742, USA} \& D.L. Shone 
			\\ 
University of Manchester, Nuffield Radio Astronomy Laboratories,
Jodrell~Bank, Macclesfield, Cheshire SK11~9DL, UK\\
$^{\star}$Present address: Department of Astronomy, University of Maryland, College Park, MD~20742, USA.}

\date{Accepted for publication in MNRAS}
\begin{document}

\maketitle			

\begin{abstract}
\large

{We present sensitive, high resolution observations of neutral
hydrogen (HI) in the unusually gas-rich, oval distortion of the
Seyfert galaxy NGC4151. The gas dynamics of the oval are found to be
consistent with those of a kinematically weak bar, fully confirming
previous suggestions for the presence of a fat bar, and, for the first
time, individual gaseous features in the bar are spatially
resolved. In particular, the two bright regions close to the leading
edges of the bar in NGC4151 exhibit kinematics strikingly similar to
the signature of bar shocks seen in gas-dynamical simulations, and
demonstrate how strong the gaseous response may be even in such a weak
bar potential. The residual velocity field, showing deviations from
circular motion, is largely consistent with streaming in a bar
potential, and, in addition, clearly shows that inflow is concentrated
in narrow regions originating in the shocks. This inflow may represent
an early stage in the fuelling process of the AGN.

The presence and properties of the shocks in NGC4151 indicate that, in
addition to the $x_1$ orbits, the family of $x_2$ orbits exist and are
of significant extent in the bar of NGC4151, with gas streaming from
the shocks making the transition between the two families. We
therefore suggest that the circumnuclear ellipse, identified optically
by previous authors and associated with gas flowing in $x_2$ orbits,
has formed as a natural consequence of the gas flows in the bar
without the requirement for a second, inner bar.

Observations of HI were previously thought to be ill-suited to the
study of bar shocks due to limitations in angular resolution and
sensitivity. However, our observations show that, following recent
instrumental enhancements, such measurements are now feasible, albeit
at the limits of instrumental capability.}

\end{abstract}

\normalsize

\section{Introduction}
 
Seyfert galaxies are the closest and most common class of galaxies
containing Active Galactic Nuclei (AGN) and, although their active nuclei
are relatively weak, they exhibit many of the properties of their more
luminous counterparts, particularly quasars. As such they represent
ideal sites for the study of the AGN phenomenon and its relationship
to the host galaxy environment, which is difficult to observe in more
distant and powerful AGN.

Observations of the $\lambda$21-cm spectral line of neutral hydrogen
(HI) provide valuable information on gas kinematics in nearby
galaxies, and may allow us to investigate models for AGN and their
hosts on a wide range of scales, from the outer-most regions, where
the gas may be affected by tidal interactions, down to the
circumnuclear regions, where it may play a role in the fuelling of
AGN. In particular, {\em bars} - which are seen in nearly 70\% of
galaxies when studied in the near IR (Mulchaey \& Regan, 1997)- may be
an efficient mechanism for transporting gas from the outer parts of a
galaxy towards the active nucleus, where other transport processes
become important (Roberts, Van Albada \& Huntley, 1979; Shlosman,
Begelman \& Frank, 1990; Larson 1994). Since neutral gas may respond
in a highly non-linear way to even small deviations from axial
symmetry, it is an excellent tracer of the underlying gravitational
potential of a barred galaxy (Teuben et al., 1986), but limitations in
angular resolution and sensitivity have, until now, prevented detailed
studies of HI in the bars of AGN hosts.

The `fuelling' of an AGN requires gas from the outer regions of a
galaxy to be delivered to its centre with essentially zero angular
momentum, in order to form (and refuel) an accretion disc around a
central black hole (Gunn, 1979; Shlosman et al., 1990). Although
controversial, gas streaming in galactic bars may play an important
role in the early stages of this process (Simkin, Su \& Schwarz,
1980), with significant angular momentum loss occurring when two
different families of gas orbits meet and form dissipative shocks,
allowing gas to move inward (Prendergast, 1983; Athanassoula,
1992a,b).  Until now, direct evidence for this has been difficult to
obtain (Lindblad \& J\"orsater, 1988; Sellwood \& Wilkinson, 1993;
Teuben, 1996) since optical studies are restricted by the small
amounts of ionised gas in bar shocks (Lindblad et al., 1996), and
detailed studies of more ubiquitous neutral hydrogen in the bars of
AGN hosts, and indeed, normal galaxies, have been limited by
inadequate angular resolution and sensitivity.

Although the physical conditions in the host galaxy of an AGN may be
intimately related to the nuclear activity, to date there have been
relatively few detailed synthesis studies of HI in AGN hosts. As part
of an ongoing project to study the distribution and kinematics of HI
in AGN (Seyferts) we present here, the highest angular resolution
observations of the neutral hydrogen in the bar of the archetypal
Seyfert galaxy NGC4151, obtained using the VLA in B configuration.

NGC4151 is a well-studied nearby gas-rich spiral galaxy with a central
oval distortion (Bosma, Ekers, Lequeux, 1977) and a Seyfert 1.5
nucleus (Osterbrock \& Koski, 1976). The oval distortion, identified
kinematically in early HI studies (Bosma et al., 1977, Bosma, 1981)
and evident in deep optical exposures (Arp, 1977) and subsequent HI
maps (Pedlar et al., 1992), is elongated along PA $\sim$130$^o$ and
has dimensions of $3.3' \times 2.1'$ (12.9 kpc $\times$ 8.2 kpc).  In
this paper, we establish the oval distortion as a weak, 'fat' bar and
concentrate particularly on the gas dynamics in this region; we relate
our study of the gas flows in the bar to models of gas flows in
non-axisymmetric barred potentials.  Assuming a heliocentric velocity
of 998 km s$^{-1}$, and H$_0$ = 75 km s$^{-1}$ Mpc$^{-1}$ , the
distance to NGC4151 is 13.3 Mpc (see also Mundell et al., 1998), so 1"
corresponds to 65pc in the galaxy.

\section{Observations and Reduction}

Full details of the observations and subsequent data processing
techniques (including production of the moment maps) are presented in
Mundell et al. (1998).  The data used in this paper were obtained from
a 63-channel, naturally weighted spectral line data cube which has an
angular resolution of 6$'' \times$ 5$''$, a spectral resolution of
10.3 km s$^{-1}$ and a greater sensitivity (lowest detectable column
density of 1.32 x 10$^{20}$ cm$^{-2}$) than the higher resolution,
uniformly weighted cube.

\section{Results}

\subsection{The ``Fat'' Bar of NGC4151}
 
NGC4151 shares some of the basic characteristics of barred spirals as
summarised by Roberts et al. (1979); for example the ``lack of
dominance of the bar in photometric studies'', which is indeed the
case for NGC4151 where the underlying mass distribution is only a mild
oval distortion (Bosma et al., 1977; Arp, 1977); the ``sharp bend of
the bar into spiral arms'' which is evident in the HI image of NGC4151
(Mundell et al., 1998), and ``gas streaming along the bar'' (see
Section 3). However, the amount of neutral hydrogen present in the
oval region (central $3.3' \times 2.1'$) of NGC4151 is unusually high
compared with other early type barred spirals (e.g. Roberts et al.,
1979; Regan et al., 1996).

Optical photometry of the outer spiral structure (Simkin, 1975) found
a major axis position angle of $\sim$26$^{\circ}$ and an inclination
of 21$^{\circ}$. Similarly, Pedlar et al., reported that the overall
HI velocity field of NGC4151 was approximately consistent with
circular rotation and an inclination of 21$^{\circ}$ for the galaxy; a
change in position angle of the line of nodes in the oval region,
however, indicated the presence of non-circular motions. Their
observations also suggested structure in the oval but were limited by
angular resolution ($\sim17'' \times 22''$) and the strong central
absorption feature. They detected regions of high HI concentration
(N$_H$ $\sim$1.5 $\times$ 10$^{21}$ cm$^{-2}$) in the north-west and
south-east regions of the bar which appeared to be small spiral arms,
with their concave edges facing toward the major axis of the bar (see
Section 3.2).

\subsubsection{Establishing the Central Oval Distortion as a Bar}

If a uniform circular disc is viewed at some inclination, i, to the
line of sight it appears as an `oval', with a semi-major axis, a, and
semi-minor axis of (b = a cos i). If the disc is rotating, the tilting
of the disc will result in an observed radial velocity field where the
zero velocity (or line of constant/systemic velocity or kinematic
minor axis) lies along the spatial minor axis, b.

In NGC4151, the central region is also oval but the iso-velocity
contours are very different compared to those expected from an
inclined circularly rotating disc. The iso-velocity contours are
elongated {\em along} the oval, indicating the presence of significant
deviations from circular motion.  Prendergast (1983) points out that,
in barred galaxies, the gas response leads the bar and so the larger
the angle between the gas (or kinematic) major axis and the bar
(spatial) major axis, the weaker the bar. For NGC4151, assuming the
bar lies in the plane of the galaxy and taking the average PA of the
line of nodes in the bar to be 28$^{\circ}$ (Mundell et al., 1998),
this angle is $\sim$72$^{\circ}$, (i.e., the kinematic major axis lies
almost on the spatial minor axis), indicating that the bar is {\em
kinematically weak}, as may be expected from its oval shape.

\subsection{Shocks in the Bar}

\begin{figure*}
\setlength{\unitlength}{1mm}
\begin{picture}(50,150)
 
\put(0,0){\includegraphics{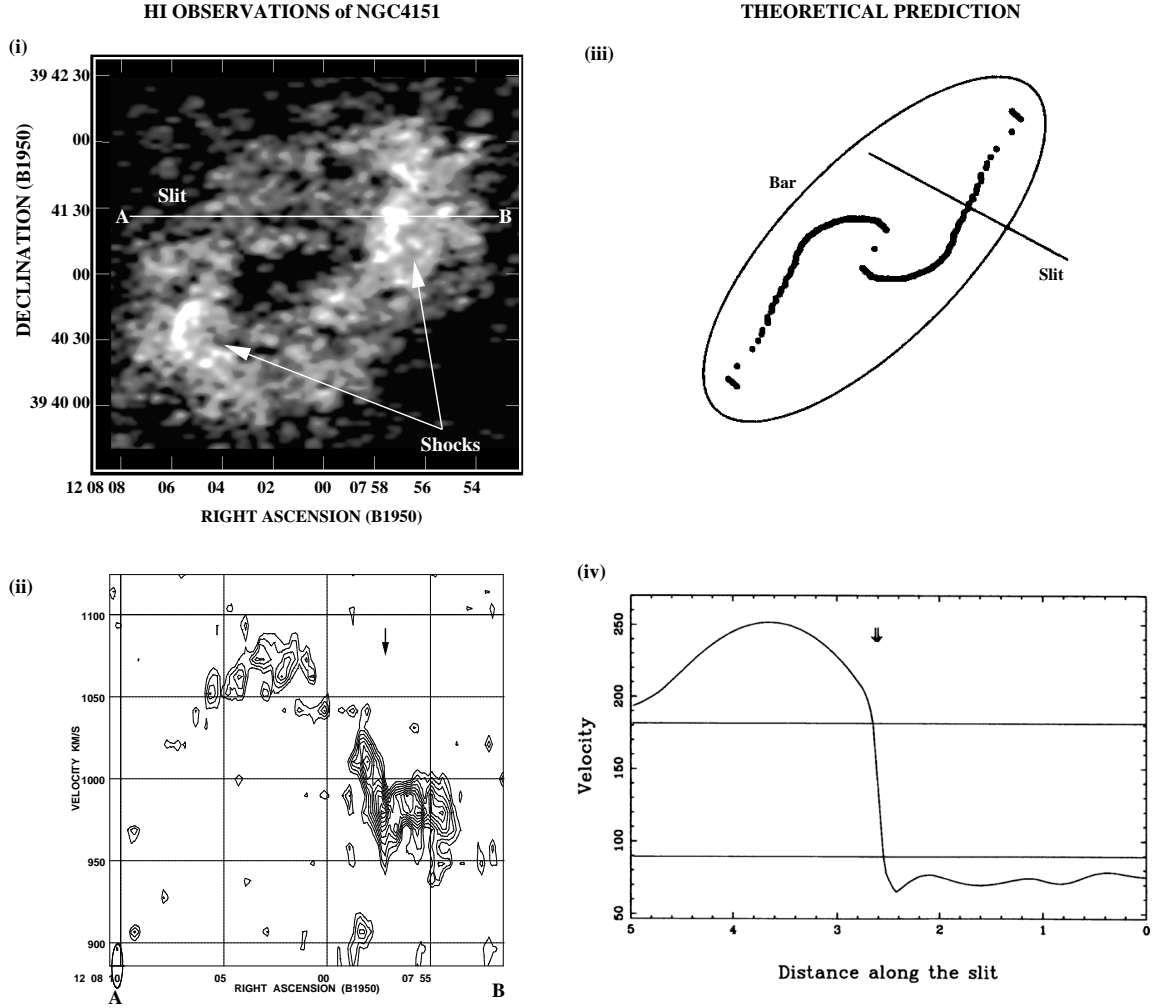}}
\end{picture} 
\caption{(a) HI emission from the bar in NGC4151. The shock regions
are indicated and the position of the slit (A -- B) from which the
spectrum was extracted is marked; (b) The change in velocity along the
length of the slit (shown in (a)). Note the sharp jump in velocity
(marked $\downarrow$) which corresponds to the bright emission region
in the NW shock; (c) Schematic from simulations of gas flows in bars
by Athanassoula (1992b) showing the gas flowing in from the $x_1$
orbits down to the smaller, lower energy, $x_2$ orbits.  The long-slit
position is marked, placed across the shock region in the simulations;
(d) The corresponding predicted velocity profile along the long-slit
(shown in (c)) derived from simulations (Athanassoula, 1992b). Note
the predicted velocity jump across the shock (marked $\downarrow$),
similar to the observed feature in NGC4151 (b).}

\label{shocks}
\end{figure*}

The weak, fat bar in NGC4151 possesses two small regions of enhanced
emission which lie in the NW and SE corners of the bar (Figure
\ref{shocks}). These two bright regions are short, curved and offset
from the major axis of the bar, i.e. they lie towards the leading
edges of the bar and are associated with the small strings of HII regions
(Perez-Fournon \& Wilson, 1990) which are visible as ''ansae'' in
optical images of the bar.  They closely resemble structures simulated
by Athanassoula (1992b), who carried out a detailed theoretical study
of the properties of periodic orbits in a wide range of barred galaxy
model potentials, in order to reproduce and explain the morphologies
and kinematics of gas observed in barred galaxies.

In particular, she applies her simulations to strongly barred galaxies
exhibiting dust lanes (such as NGC1300 and 6782) and also examines the
behaviour of gas in weak, oval bars.  Leading dust lanes in  bars
are typical of early to intermediate Hubble type barred spirals, and
are especially characteristic of type SBb (Sandage 1961; Sandage and
Bedke 1994).  Prendergast (1962, unpublished - Sellwood
\& Wilkinson, 1993) associated these dust lanes with shocks in the gas
flows and subsequently some circumstantial evidence has been
discovered to support this idea.  For example some dust lanes are
particularly prominent in radio continuum observations because the
shocks are regions of compressed gas, dust, magnetic fields and high
energy particles - an environment ripe for the production of radio
continuum emission (Ondrechen \& van der Hulst, 1983; Ondrechen,
1985). In fact, large scale (kpc) extended radio continuum emission
has been detected in NGC4151 (Baum et al., 1993). At $\lambda$21-cm
this weak extended continuum shows some correspondence with the bright
HI emission arcs, although Baum et al. (1993) suggest a
starburst-driven superwind for its origin.

A simple and direct way to determine whether shocks are present in the
gas in NGC4151 is to measure the velocity field across each arc and
look for the predicted velocity jumps that are characteristic of
shocks in galactic (bar) gas. Unfortunately, previous {\em optical}
studies of the bar have failed to reveal evidence for shocks; B-I
colour maps (Vila-Vilaro et al., 1995) show only a circumnuclear dust
ring on scales much smaller (only 11''$\times$ 18'') than the HI
shocks. In addition, detailed long-slit spectroscopy of the small
string of HII regions, observed towards each end of the bar, have
shown that the ionised gas is typical of HII regions photoionised by
young star clusters (Schulz, 1985); although the kinematics of the
ionised gas show both inward and outward radial motions of $\sim$20 km
s$^{-1}$ in addition to a circular component, these are thought to be
merely a natural consequence of gas flow in elliptical orbits
(Schultz, 1985). Neutral hydrogen, however, should be ideal for this
type of measurement but until now has proved useless due to
limitations in angular resolution and sensitivity. Since the
concentration of {\em HI} in the bar of NGC4151 is high, especially in
the shock regions, the high resolution of our new HI observations
permit detailed investigation of the gas kinematics in the bar of
NGC4151, making it is possible to search for these velocity jumps that
are characteristic of shocks in the neutral gas.

We examined emission from the HI cube along the equivalent of a
6$''$-wide slit, at several locations and orientations, across the
length of each emission arc. When the slit is positioned perpendicular
to the emission arc and across its brightest peak (Figure
\ref{shocks}) the maximum velocity jump is seen, and an example of
this for the NW arc is shown in Figure \ref{shocks}. Similarly
oriented slits centred at different positions along both arcs showed a
similar velocity structure. These observations bear a striking
resemblance to the theoretical predictions of gas behaviour in a weak
barred potential (Athanassoula, 1992b), also shown in Figure
\ref{shocks}, in which the maximum velocity jump in the simulated data
is also found when the slit is perpendicular to the shock. It should
be noted that the simulations are for a general weak bar and were {\em
not} intended to simulate the conditions in NGC4151; consequently the
similarity between observation and theory seems even more striking.

\subsubsection{Inflow Along the Shocks}

\begin{figure*}
\setlength{\unitlength}{1mm}
\begin{picture}(50,100)
 
\put(0,0){\includegraphics{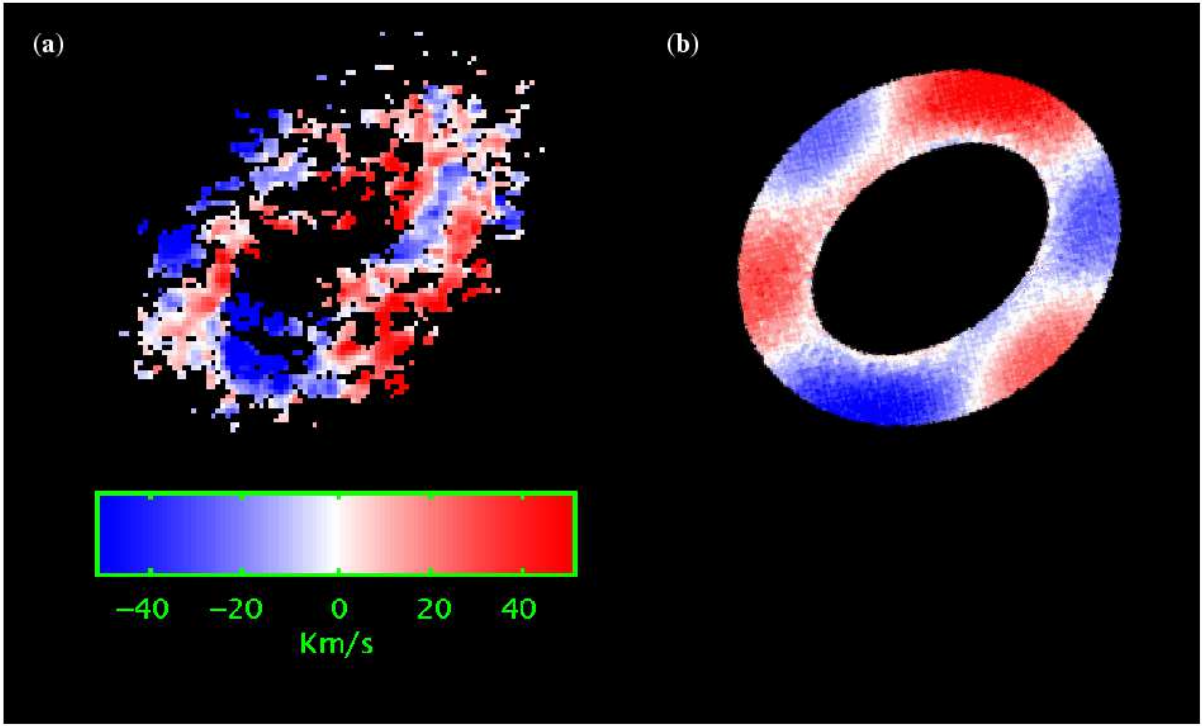}}
\end{picture} 
\caption{(a) The residual velocity field of the bar in NGC4151, with
the circular component of motion removed; (b) Residual velocity field,
after subtraction of fitted circular motion (in the same manner as for
{\bf Fig. 2.}a), for NEMO N-body simulation of 20,000 particles in
barred potential, viewed in the same orientation as NGC4151.  The
potential is chosen to be qualitatively similar to NGC4151.  Both
images are colour coded to represent streaming motions which deviate
from circular motion (white); red-tinted regions are receding faster
than circular motion about the systemic velocity, blue regions are
slower than systemic.  Thus, given the orientation of NGC4151, the
red-shifted region originating in the south-eastern shock and the blue
shifted counterpart in the north-west indicate inflow.}
 
\label{resids}
\end{figure*}

The iso-velocity contour maps (``spider diagrams'') often employed to
display velocity fields reveal deviations from circular motion, but
are often inadequate for distinguishing between the radial component
of motion in a barred potential and true inflow due to loss of angular
momentum in shocks.  Therefore, in order to investigate the bar
kinematics in NGC4151 more thoroughly, we removed a circular
rotational component (as derived in Mundell et al., 1998) from the
observed velocity field to produce the residual velocity field shown
in Figure \ref{resids}(a).  Deviations from circular motion are
expected for non-circular orbits in an ellipsoidal bar potential, with
equal amounts of inflow and outflow.  However, the shock regions are
strikingly apparent in this image, displaying strong deviation from
the local bar field. Assuming that the large-scale spiral arms are
trailing, implying clockwise rotation of NGC4151, the galactic disc is
tilted such that the north-west half of the disc is furthest from us.
The bar is thought to lie in the plane of the galactic disc, and so
the motion of gas in the shocks corresponds to inflow along the bar.

For comparison, we have produced a simple model for the velocity field
of a bar which is qualitatively similar to that in NGC4151 observed at
the same orientation.  The model, produced using the NEMO (Teuben, 1985)
stellar dynamics toolbox contains particles only, so that it is
indicative of orbits in a barred potential, but does not show more
complex gas dynamical effects such as shocks.  The distinction between
the behaviour of gas and stars is important; since stars do not behave
as a fluid, stellar orbits may cross, whereas crowding of gas
streamlines corresponding to these orbits may lead to non-linear
phenomena such as shocks.  The residual velocity field produced by our
simulation is intended to show the deviations from circular motion
expected for the elliptical stellar orbits of a barred potential
(Figure \ref{resids}(b)). These deviations appear as inflow and
outflow, but there is no {\em net} flow either way in this
non-dissipative model. Thus it should be noted that observations of
streaming motions in galactic bars do not necessarily imply net
inflow; these may simply be the radial components of non-circular
motion expected in a non-axisymmetric bar potential.

Sophisticated simulations which {\em do} include gas dynamical effects
have been conducted by several authors (e.g. Sanders \& Tubbs, 1980;
Van Albada \& Roberts, 1981; Schwarz, 1981, 1984; Combes \& Gerin,
1985; Athanassoula, 1992a,b; Sellwood \& Wilkinson, 1993, and
references therein) to explain features, such as shocks, in barred
galaxies in terms of the interaction between various families of
periodic orbits in the bar. The existence of such orbits then allow
various physical parameters of the bar, such as the presence of
possible resonances, to be inferred.

\section{Discussion}

\subsection{Periodic Orbit Families, Shocks and Resonances}
 
There exist a large number of studies of orbits in bar potentials
(Contopoulos \& Grosb$\o$l, 1989 and references therein), although not
all orbit patterns produced in simulations are relevant to the
structure of real bars (Sellwood \& Wilkinson, 1993). The two main
families of periodic orbits (using the notation of Contopoulos \&
Papayannopoulos, 1980) are the $x_1$ and $x_2$ families. The $x_1$
family are elongated along the bar major axis and support the bar,
while the $x_2$ are elongated perpendicular to the bar. The $x_3$
family are also perpendicular to the bar but are more elongated than
the $x_2$ family and are unstable (Sellwood \& Wilkinson, 1993).
 
The presence and properties of the different orbit families depend on
the existence and location of resonances within the bar. The $x_1$
orbits exist within the co-rotation radius (CR). If no inner Lindblad
resonance (ILR) exists then no other families exist and the bar is
populated by $x_1$ orbits only. If one or more ILRs exist in the bar
then $x_2$ orbits are present and lie within the first (outermost)
ILR; if there are two ILRs the $x_2$ orbits lie between the two ILRs
(Athanassoula, 1992a, Shaw et al., 1993).  The angular velocity of the
bar, the global mass distribution and the degree of central
concentration, in turn, determine the possible existence and locations
of resonances in the bar (Teuben et al., 1986). For example, fast
rotating bars with low central mass concentrations have no ILRs (Shaw
et al., 1993).

Our understanding of gas flows in barred potentials has been
significantly increased by such theoretical studies of orbit families and
detailed numerical simulations, but direct observations of predicted
phenomena, in particular kinematic signatures, provide valuable
verification of such theories.

As described in Section 3.2.1, stars do not behave as a fluid so
stellar orbits may cross, whereas the crowding of gas streamlines
corresponding to these orbits may lead to non-linear phenomena such as
shocks. Shocks associated with the crowding of streamlines may be
produced when $x_1$ orbits, alone, are present in a bar, but only when
the bar potential is strong (e.g., straight, narrow dust lanes
bisecting a galaxy). However, Athanssoula (1992b) has shown that in
order for the shock loci to be short, curved {\em and} offset from the
bar major axis, as in NGC4151, the family of $x_2$ orbits must also
exist and be of {\em large} extent. This occurs when the bar potential
is weak, has a low quadrupole moment, a high central concentration and
a pattern speed such that the co-rotation radius lies in the range
1.2$\pm$0.1a (a is the semi-major axis of the bar).  The $x_1$ orbits
also have high curvatures near their apocentres in the regions in
which the shocks occur in order for the shock loci to form along the
leading edge of the weak bar.  The shape of the shocks can then be
explained by a gradual shift in the orientation of the flow lines from
along the bar ($x_1$ orbits) to perpendicular to the bar. Angular
momentum is dissipated in the shock region and the gas then quickly
settles onto smaller, lower energy, $x_2$ orbits, closer to the
nucleus. Our observations of the shocks in NGC4151 provide compelling
evidence for the presence of these two orbit families in the bar, with
gas streaming from the shock regions making the transition between the
two families. In fact, recent optical studies (Vila-Vilaro, et al.,
1995; Asif et al., 1998) have found evidence for the presence of a
circumnuclear ellipse of dust and gas ($11'' \times 18''$), elongated
perpendicular to the bar. Such a ring is thought to form as a result
of gas flowing in $x_2$ orbits (Shlosman, 1996), interior to the outer
ILR, and in fact, each of the inner ends of the inflow we observe lie
close to the ends of the major axis of this ellipse, providing clear
evidence for such a transition. The circumnuclear ellipse in NGC4151
may therefore have formed as a natural consequence of the non-linear
gas dynamics in the bar and does not require the existence of an inner
stellar bar. This is consistent with the results of IR studies in
which no inner bar is detected down to the central 3$"$
(Alonso-Herrero, et al., 1998).

\subsection{Bar Evolution, Inflow and Fuelling the AGN}
 
In a non-axisymmetric barred potential, gas flow has a complicated
structure, with regions of both outflow and inflow (Athanassoula,
1995). If there are no shocks in the bar, the gas follows
quasi-elliptical orbits and there is only a small net inflow due to
viscosity. If, however, shocks are present, the gas changes direction
and moves inwards at high speeds after hitting the shock. The average
inflow velocity however is only a few km s$^{-1}$ despite local inflow
velocities as high as 100 km s$^{-1}$ (Athanassoula, 1992b).  It is
thought that it is also easier for gas to flow into the centre when no
ILRs are present.
 
Simulations have also shown that two categories of inflow occur
depending on the stage of evolution of the bar; high inflow rates
occur during the bar formation stage and much smaller inflow rates
occur once the bar reaches a quasi-steady state (Athanassoula, 1994).

When the bar first begins to form, no central concentration or ILRs
are present. Gas inflow at this early stage is very efficient and gas
is pushed towards the centre. As time passes, the central mass builds
up and the bar pattern speed decreases leading to the formation of
ILRs and the quasi-steady-state with smaller inflow rates. This type
of negative feedback loop would seems rather unpromising for the
prospect of fuelling an AGN. However, the presence of ILRs, central
concentration, and slow pattern speed leads to the formation of
leading edge shocks which in turn cause inflow, albeit at a smaller
rate than the initial bar formation stage. The streaming velocities
observed in NGC4151 of up to 40 km s$^{-1}$ are consistent with those
predicted by simulations (Athanassoula, 1992b).
 
This less dramatic type of fuelling mechanism may in fact be ideal for
Seyferts. Statistical evidence shows that strongly interacting
galaxies and highly disturbed systems do not show an excess of Seyfert
activity (Bushouse, 1986). This suggests that Seyfert activity is
strongly related to the host galaxy properties and that nuclear
activity may be the result of the galaxy responding coherently with
well-structured features to non-axisymmetric perturbations of the
potential rather than due to major disruption (Moles, Marquez \&
Perez, 1995).

\section{Conclusions}

Spectral imaging of HI in NGC4151 has shown that the central oval
distortion exhibits the kinematic characteristics of a weak bar, and
in addition, bright regions close to the leading edges show sharp
changes in the velocity field strikingly similar to shocks in
gas-dynamical simulations by Athanassoula.  This implies the presence
of both the $x_1$ and $x_2$ orbit families, since the offset shocks in
weak bars are due to convergence of orbital streamlines from the two
families.
 
The residual velocity field shows streaming consistent with a weak bar
potential, but with gas inflow clearly directed along narrow channels
originating in the shocks and leading to the inner dusty ellipse seen
optically by Vila-Vilaro {\it et al}.  We identify this ellipse with
the lower energy $x_2$ orbits, and suggest that it has formed as a
consequence of the gas flows in the bar, without the requirement for a
second, inner bar. The inflow along the bar may represent an early
stage in the fuelling of the AGN.

These HI observations of NGC4151 demonstrate that it is now possible
to study the detailed kinematics of gas closer to the nuclear regions
than has hitherto been thought possible, and confirm long-standing
theoretical predictions of gas inflow in bars.  Further such studies
of bars in nearby active and normal galaxies, and comparison with detailed
simulations may provide the opportunity to fully understand this
potentially important stage of the nuclear fuelling chain.         

\section{Acknowledgements}
We thank Alan Pedlar, Lia Athanassoula and Elias Brinks for helpful
discussions. We are grateful to the referee, Professor Ron Buta, for
helpful comments which improved an earlier draft of this paper.  CGM
acknowledges a research studentship and research fellowship from the
U.K. Particle Physics and Astronomy Research Council (PPARC).\\

{}

\end{document}